\documentclass[lettersize,journal]{IEEEtran}
\usepackage{amsmath,amsfonts}
\usepackage{algorithmic}
\usepackage{algorithm}
\usepackage{array}
\usepackage[caption=false,font=normalsize,labelfont=sf,textfont=sf]{subfig}
\usepackage{textcomp}
\usepackage{stfloats}
\usepackage{url}
\usepackage{verbatim}
\usepackage{graphicx}
\usepackage{cite}
\usepackage{bm}
\usepackage{booktabs}
\usepackage{tabularx}
\usepackage{multirow}
\usepackage{balance}
\usepackage{orcidlink}
\usepackage{xcolor}

\begin{document}

\title{Heterogeneous Mapping for Analog In-Memory Computing Accelerators: A Unified Workflow}

\author{Corey Lammie\orcidlink{0000-0001-5564-1356},~\IEEEmembership{Member,~IEEE,}\thanks{\hspace{-1em}\rule{3cm}{0.5pt} \newline \textcopyright  \hspace{1pt} 2026 IEEE. Personal use of this material is permitted. Permission from IEEE must be obtained for all other uses, in any current or future media, including reprinting/republishing this material for advertising or promotional purposes, creating new collective works, for resale or redistribution to servers or lists, or reuse of any copyrighted component of this work in other works.}

\thanks{Corey Lammie is with IBM Research -- Zurich, Switzerland (e-mail: corey.lammie@ibm.com).}}

\markboth{IEEE Computer Architecture Letters}%
{Lammie: Heterogeneous Mapping for AIMC Accelerators: A Unified Workflow}

\maketitle
\begin{abstract}
Analog In-Memory Computing (AIMC) accelerators execute matrix--vector multiplications directly within memory arrays, reducing data movement and improving DNN inference efficiency. Their limited effective precision motivates heterogeneous architectures that combine analog compute tiles with digital processing units. This letter classifies existing methods for partitioning DNN workloads across these resources by mapping granularity, optimization strategy, and model support, and distills them into a unified four-stage workflow. To demonstrate the workflow on a model class not yet addressed by existing methods, we apply its first two stages to GPT-2, producing the first AIMC-specific precision sensitivity profile for a decoder-only transformer. Sensitivity is dominated by 4 of 49 projections, with the first decoder block's attention output dominating by an order of magnitude. This suggests that projection-level mapping and selective digital execution of early-block and output-facing projections are important for reliable decoder-transformer deployment on AIMC hardware.
\end{abstract}


\section{Introduction}
\IEEEPARstart{T}{he} growing computational demands of DNN inference have driven the development of domain-specific accelerator architectures that overcome the data movement bottleneck of conventional von Neumann systems. Analog In-Memory Computing (AIMC) is a prominent in-memory processing paradigm that executes weight-stationary matrix--vector multiplications (MVMs) directly within memory arrays, substantially reducing off-chip data transfers and improving energy efficiency~\cite{Sebastian2020}. AIMC-based accelerators have been demonstrated at scale, with multi-core mixed-signal chips.

However, the reduced effective precision of analog compute tiles (typically 3--8 bits due to device variability and peripheral circuit quantization) limits their applicability to precision-sensitive operations~\cite{Rasch2023}. This has motivated \textit{heterogeneous accelerator architectures} that integrate AIMC tiles alongside digital Compute Units (CUs): analog tiles handle precision-tolerant MVMs, while digital CUs execute sensitive layers or non-linear operations at full precision~\cite{Ueyoshi2022}.

\begin{figure}[!t]
	\centering
	\includegraphics[width=0.48\textwidth]{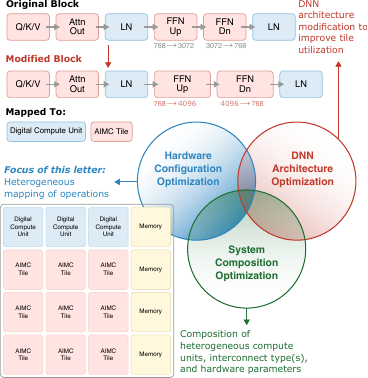}
	\caption{Overview of co-optimization strategies for AIMC-based accelerators. Heterogeneous mapping of DNN operations to analog and digital compute units (the focus of this letter) lies at the intersection of hardware configuration, DNN architecture, and system-level optimization.}
	\label{fig:overview}
\end{figure}

The central \textit{mapping problem}, assigning DNN operations to analog or digital CUs to jointly optimize accuracy, energy, and latency, is combinatorial ($K^L$ mappings for $K$ precision levels and $L$ layers) and multi-objective, making exhaustive search intractable for practical models.
As illustrated in Fig.~\ref{fig:overview}, heterogeneous mapping is not an isolated design step: it couples hardware configuration, DNN architecture, and system-level optimization, and is therefore a workload scheduling problem over heterogeneous compute resources.

Despite growing interest, existing mapping methods are fragmented: most target only CNNs, lack a standardized evaluation methodology, and rarely quantify their own computational cost. This letter makes three contributions: (i)~a classification of existing mapping methods across granularity, strategy, and architecture support; (ii)~a unified end-to-end mapping workflow distilled from these methods; and (iii)~the first AIMC-specific precision sensitivity analysis of a decoder-based transformer (GPT-2), extending heterogeneous mapping beyond CNNs and encoder-based transformers covered by prior work.

\section{Existing Mapping Methods}

Table~\ref{tab:comparison} compares nine heterogeneous mapping methods along six dimensions. We organize the discussion around three axes that most strongly differentiate them.

\begin{table*}[!t]
\centering
\caption{Heterogeneous mapping methods for AIMC accelerators. Gran.=Granularity; FP=floating-point; HWA=hardware-aware trained; QAT=quantization-aware training; Eval.=Evaluation platform; Cost=computational cost of the mapping method; Enc.\ Tr.=Encoder-based Transformers.}
\label{tab:comparison}
\footnotesize
\setlength{\tabcolsep}{3pt}
\begin{tabularx}{\linewidth}{lllXlll}
\toprule
\textbf{Method}   & \textbf{Gran.} & \textbf{Init.} & \textbf{Strategy}  & \textbf{Architectures}  & \textbf{Eval.}  & \textbf{Cost}  \\
\midrule
PAWDD~\cite{Kao2022}   & Layer  & FP  & Empirical precision categorization of layers.  & CNNs & Synth.\ 7nm  & Low \\
CIMQ~\cite{Bai2024}    & Array  & FP  & Sequential layer partitioning with channel reordering.  & CNNs & NeuroSim~\cite{Peng2019}  & Med. \\
OSA-HCIM~\cite{Chen2024} & Array  & HWA  & Runtime saliency evaluation for analog--digital ratio.  & CNNs & Synth.\ 65nm  & Med. \\
Harmonica~\cite{Behnam2024} & Chan.  & HWA  & Per-channel Hessian eigenpair analysis.  & CNNs & Sys-lvl~\cite{Shafiee2016}  & High \\
Hessian~\cite{Dash2022} & Wt.  & HWA  & Per-weight Hessian-based precision allocation.  & CNNs & NeuroSim~\cite{Peng2019}  & High \\
ODiMO~\cite{Risso2025} & Chan.  & FP  & Differentiable cost-aware opt.\ during training. & CNNs & DIANA~\cite{Ueyoshi2022}  & High \\
LionHeart~\cite{Lammie2025} & Layer  & FP  & Heuristic layer-wise retraining and assignment. & CNNs, Enc.\ Tr. & ALPINE~\cite{Klein2023}  & Med. \\
RH-IMC~\cite{Chen2025} & Input  & FP  & MSB/LSB bit-split across digital/analog units.  & CNNs, RNNs & RTL emu.  & Low \\
MPS~\cite{Benmeziane2026} & Layer  & FP  & Supernetwork with fairness training \& Pareto-rank opt. & CNNs, ViTs, BERT & 3DCiM~\cite{3DCiM}  & Med. \\
\bottomrule
\end{tabularx}
\end{table*}

Mapping \textit{granularity} ranges from entire layers or layer groups (PAWDD, LionHeart, MPS), analogous to coarse-grained task scheduling, to individual weights (Hessian-driven~\cite{Dash2022}), representing fine-grained resource allocation. Coarser granularity simplifies dataflow and reduces inter-unit communication, but limits optimization freedom; finer granularity can improve accuracy--energy trade-offs at the cost of increased interconnect traffic and system complexity.

Mapping \textit{strategies} fall into three categories. Heuristic methods (PAWDD, LionHeart) apply rules or iterative retraining and are fast but may miss optimal solutions; the widely used \emph{first-last} scheme, which maps the first and last layers to digital CUs, is the simplest example. Analytical methods (Harmonica, Hessian-driven) use second-order sensitivity analysis for mathematically grounded allocation, but Hessian computation scales poorly beyond ResNet-50. Learning-based methods embed the mapping decision into training itself: ODiMO~\cite{Risso2025} uses differentiable relaxation, while MPS~\cite{Benmeziane2026} constructs a mixed-precision supernetwork with analog, INT8, and FP16 paths per layer, using gradient-based Pareto optimization to jointly search over precision and mapping.

Methods also differ in their starting point: some operate on original FP weights (PAWDD, CIMQ, ODiMO, LionHeart, MPS), while others start from hardware-aware trained (HWA) weights and selectively promote sensitive components to higher-precision CUs (OSA-HCIM, Harmonica, Hessian-driven). The HWA path can yield higher accuracy but adds an upfront training cost.
A critical gap is workload coverage. All methods support CNNs, but only LionHeart and MPS handle encoder-based transformer architectures, only RH-IMC supports RNNs, and only MPS covers vision Transformers and BERT. 
No method explicitly targets decoder-based LLMs. Autoregressive execution, KV-cache-dependent memory behavior, and repeated deep decoder blocks do not fundamentally preclude extension of prior mapping methods, but they introduce additional modeling and evaluation challenges for heterogeneous mappings.

Digital mixed-precision methods (e.g., GPTQ, AWQ) perform per-layer sensitivity analyses for Transformers but model only uniform quantization noise, not the compound analog non-idealities (device variability, crossbar tiling, ADC/DAC effects) that dominate in AIMC. Evaluation platforms also vary widely, making direct cross-method comparison of energy and latency figures unreliable.

\section{Unified Mapping Workflow}

\begin{figure}[!t]
	\centering
	\includegraphics[width=0.48\textwidth]{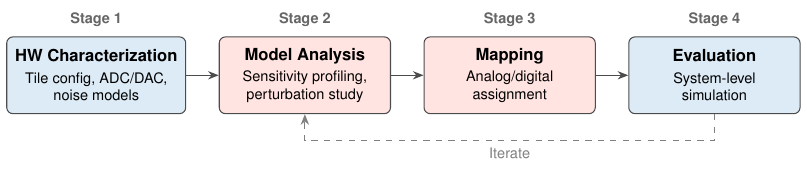}
	\caption{Unified heterogeneous mapping workflow. Stages iterate until accuracy--energy constraints are satisfied.}
	\label{fig:workflow}
\end{figure}

From the methods in Section~II we distill a four-stage workflow (Fig.~\ref{fig:workflow}) applicable to any heterogeneous AIMC accelerator.

The first stage, \textit{architecture characterization}, profiles the target accelerator: tile count and configuration, digital CU capabilities (INT8, FP16, FP32), tile I/O precision, memory hierarchy, and interconnect topology.
Candidate tiling strategies should also be evaluated in terms of array utilization, weight reuse, arithmetic intensity, and resulting memory traffic. For volatile or charge-based memories that require refresh, as well as non-volatile memories that exhibit time-dependent drift or maintenance costs, the associated latency and energy overheads should be included in the architecture model.
Surrogate models for accuracy and performance prediction are developed at this stage. 
Recent tools such as CiMLoop~\cite{Andrulis2024} provide flexible and accurate modeling support for IMC architectures and can be used as part of the architecture characterization and system-level evaluation stages of the proposed workflow.
Proxy metrics such as the \textit{analog MAC ratio} (fraction of MACs on analog tiles) and \textit{digital weight size} (on-chip memory footprint of digitally-mapped weights) are effective when validated against system-level simulation~\cite{Benmeziane2026}. For non-volatile analog memory (e.g., PCM, ReRAM), time-dependent precision degradation must also be modeled.

In the second stage, \textit{precision sensitivity analysis}, one determines which DNN components require digital execution and which tolerate analog execution. Techniques range from perturbation-based profiling ($O(L)$ forward passes) to Hessian eigenanalysis~\cite{Dash2022} and activation-based saliency~\cite{Chen2024}. This analysis prunes the mapping search space and determines the adaptation order. A common empirical observation in prior mixed-precision and heterogeneous analog/digital work is that early layers and output-facing layers often warrant higher precision, since perturbations close to the input can corrupt all downstream representations, while perturbations close to the output directly affect the final logits or prediction. The degree to which this holds remains model- and task-dependent.

The third stage, \textit{architecture-aware mapping}, applies an optimization strategy guided by the previous analysis to schedule operations across heterogeneous resources. Approaches range from heuristics to supernetworks~\cite{Benmeziane2026} that embed multiple precision paths per layer and use gradient-based multi-objective optimization. Architecture adaptations, such as adjusting layer dimensions to improve tile utilization, can be applied jointly. The output is a Pareto set of mappings trading accuracy against energy and latency.

Finally, \textit{system-level evaluation} validates the selected mapping using architecture simulators (DNN+NeuroSim~\cite{Peng2019}, ALPINE~\cite{Klein2023}, 3DCiM~\cite{3DCiM}) or physical hardware. We recommend reporting task-specific accuracy with error bars, analog MAC ratio, end-to-end latency, energy per inference, and the computational cost of the mapping search.

\section{Extension to a Decoder-Based Transformer}

As identified in Section~II, no existing heterogeneous mapping method explicitly targets or evaluates decoder-only Transformers, the architecture underlying modern LLMs. We apply Stages~1--2 of the workflow to GPT-2-small (124M parameters, 12 decoder blocks), producing the first precision sensitivity profile for a decoder-only language model under AIMC noise models. This profile is the prerequisite input that every Stage~3 method (heuristic, analytical, or learning-based) requires, and unlike Stages~3--4, which depend on a concrete accelerator configuration, it is reusable across target platforms. Completing these stages for a specific accelerator is a natural next step enabled by this analysis.

In Stage~1, we model a representative AIMC accelerator configuration. Each layer mapped to analog undergoes Gaussian weight clipping ($2.5\sigma$), tiling to $512{\times}512$ crossbar dimensions, and ${\sim}8$-bit ADC/DAC quantization. Additive Gaussian programming noise with standard deviation $\sigma_w{=}0.023$ (relative to the programmed weight range), calibrated to measured PCM conductance variability at an effective precision of ${\sim}4$~bits~\cite{AIHWKit}, is then injected. This captures the dominant noise source at $t{=}0$ (immediately after programming); time-dependent conductance drift and read noise are complementary effects that shift absolute $\Delta$ values but are expected to preserve the relative sensitivity ranking, since they scale with the same weight magnitudes. All analog conversions use the open-source AIHWKIT framework.

\begin{figure*}[!t]
	\centering
	\includegraphics[width=0.975\textwidth]{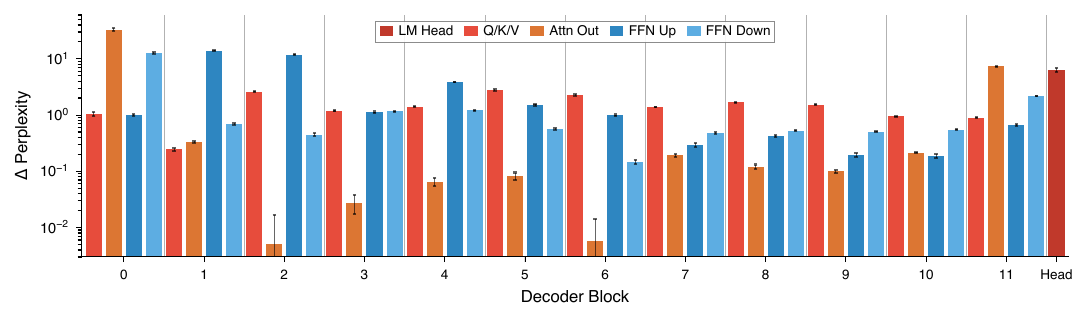}
	\caption{Per-layer precision sensitivity ($\Delta$ perplexity, mean $\pm$ std over $n{=}10$ noise realizations) for GPT-2-small. Each bar represents one weight projection, converted individually to analog while all others remain in FP32. Layers are ordered by decoder block (0--11); the final bar is the language model head.}
	\label{fig:gpt2_sensitivity}
\end{figure*}

In Stage~2, we perform a perturbation-based precision sensitivity analysis. GPT-2-small contains 49 weight projections: 12 blocks $\times$ \{fused Q/K/V~(\texttt{c\_attn}), attention output~(\texttt{c\_proj}), FFN~up~(\texttt{c\_fc}), FFN~down~(\texttt{mlp.c\_proj})\} $+$ the language model head. For each projection, we convert that single layer to analog (applying weight clipping, tiling, quantization, and programming noise) while keeping all others in FP32, and evaluate perplexity on WikiText-103 over $n{=}10$ independent noise realizations. The mean per-layer perplexity increase $\Delta_\ell = \overline{\text{PPL}}_\ell^{\,\text{analog}} - \text{PPL}^{\text{digital}}$ quantifies that layer's precision sensitivity (Fig.~\ref{fig:gpt2_sensitivity}).

The analysis reveals three structural patterns distinct from those reported for CNNs and encoder Transformers~\cite{Benmeziane2026}. First, sensitivity is dominated by a single outlier: the attention output projection (\texttt{c\_proj}) of block~0, with $\Delta{=}33.1$, an order of magnitude above most other layers. This projection is the first point at which the residual stream aggregates attention-weighted representations; noise injected here propagates through all subsequent blocks without prior residual redundancy to absorb it. The FFN~up projections of blocks~1--2 ($\Delta{=}14.0$, $11.8$) and the FFN~down of block~0 ($\Delta{=}12.8$) form a second tier. Together, these four projections account for the vast majority of total sensitivity, while 40 of 49 projections have $\Delta{<}3$, indicating that the analog precision bottleneck is highly localized rather than distributed. Second, the LM~head ($\Delta{=}6.4$) and block~11's attention output ($\Delta{=}7.3$) confirm that the last layer is precision-sensitive, consistent with the first-last pattern observed in CNNs, but the most sensitive layer overall is an \emph{internal} projection, a departure from CNN behavior. Third, attention output projections in middle blocks (2--10) are remarkably noise-tolerant ($\Delta{<}0.22$), whereas Q/K/V projections maintain moderate sensitivity ($\Delta{\approx}1$--$3$) throughout. This asymmetry within the attention datapath, where the output projection is either the most or least sensitive depending on block depth, suggests that a uniform per-block mapping is suboptimal; projection-level granularity is required. (The fused \texttt{c\_attn} projection follows GPT-2's native architecture; separating Q, K, and V as in newer Transformer implementations would enable finer-grained profiling.)

These findings directly enable Stage~3 for any target accelerator: the sensitivity profile defines which projections are candidates for analog execution, reducing the combinatorial search space from $2^{49}$ (exhaustive) to a tractable subset. Since only 9 of 49 projections have $\Delta{>}3$, a conservative threshold rule maps the remaining 40 (${\sim}82\%$ of projections) to analog tiles, providing a conservative projection-level estimate of analog coverage. For refined approaches, the profile initializes path weights in the MPS supernetwork~\cite{Benmeziane2026} or determines the retraining order in LionHeart~\cite{Lammie2025}.

One-at-a-time perturbation captures marginal sensitivity but not inter-layer interactions that may cause super-additive degradation when multiple layers operate in analog simultaneously. However, marginal sensitivity is the standard input to all surveyed Stage~3 methods; joint effects are addressed during Stage~3 optimization itself (e.g., iterative retraining or supernetwork training). The full perturbation study requires 490 forward passes ($49 \times 10$ realizations).

\section{Discussion and Open Challenges}

Scaling heterogeneous mapping to production LLMs raises open problems. Billion-parameter models introduce dynamic KV-caches that complicate memory-capacity, bandwidth, and tiling decisions for weight-stationary tile execution, asymmetric prefill/decode resource requirements, and search spaces of $2^{80} \approx 1.2 \times 10^{24}$ for a 70B model.
In particular, KV-cache storage reduces the effective memory available for weights, activations, and tiled intermediate buffers, which can change the feasible tile sizes and analog/digital partition selected by the mapping stage.

Hierarchical strategies and sub-quadratic search algorithms will be essential. GPT-2-small's 12-block pre-norm decoder architecture is structurally representative of larger models in the GPT family, and the perturbation methodology scales linearly ($O(L)$ forward passes), making extension to billion-parameter models straightforward in cost. Whether the concentration of sensitivity in early-block projections persists at scale remains an important open question. Analog memory technologies also exhibit time-dependent precision degradation (e.g., PCM conductance drift) that can invalidate a static mapping within hours, requiring runtime adaptation mechanisms. While our sensitivity profile uses a single PCM-calibrated noise model, the relative ranking of projections reflects the model's intrinsic sensitivity structure and is expected to be robust across configurations with comparable effective precision; validating this across SRAM and ReRAM noise models is future work.

From a practical standpoint, few methods quantify their own computational cost, and methods are evaluated using incompatible benchmarks and simulation tools, making fair comparison difficult. A community benchmark suite with reference accelerator configurations and calibrated precision models would significantly advance the field. Our GPT-2 results also highlight a granularity trade-off: sensitivity varies across projection types within the same block, meaning that block-level granularity, used by most existing methods, would misallocate resources, while weight-level granularity would impose prohibitive routing complexity. Projection-level mapping appears to strike a practical balance for transformers, aligning with data path boundaries already present in hardware.

\section{Conclusion}
This letter classifies heterogeneous mapping methods for AIMC accelerators, presents a unified four-stage workflow, and provides the first AIMC-specific precision sensitivity profile for a decoder-only Transformer (GPT-2). Our results motivate projection-level mapping for decoder Transformers and provide a reusable input for Stage~3 mapping methods. Key open challenges include scaling to billion-parameter LLMs, runtime adaptation for precision-degrading substrates, and standardized benchmarks.

\balance
\bibliographystyle{IEEEtran}
\bibliography{references}

@Article{Sebastian2020,
  author   = {Sebastian, Abu and others},
  journal  = {Nature Nanotechnology},
  title    = {Memory devices and applications for in-memory computing},
  year     = {2020},
  volume   = {15},
  number   = {7},
  doi      = {10.1038/s41565-020-0655-z},
}

@Article{Rasch2023,
  author   = {Rasch, Malte J. and others},
  journal  = {Nature Communications},
  title    = {Hardware-aware training for large-scale and diverse deep learning inference workloads using in-memory computing-based accelerators},
  year     = {2023},
  volume   = {14},
  number   = {1},
  doi      = {10.1038/s41467-023-40770-4},
}

@InProceedings{Ueyoshi2022,
  author    = {Ueyoshi, Kodai and others},
  booktitle = {IEEE Int. Solid-State Circuits Conf.},
  title     = {{DIANA}: An end-to-end energy-efficient digital and analog hybrid neural network {SoC}},
  year      = {2022},
  volume    = {65},
  doi       = {10.1109/ISSCC42614.2022.9731716},
}

@Article{Benmeziane2026,
  author   = {Benmeziane, Hadjer and others},
  journal  = {Nature Communications},
  title    = {Supernetwork-based efficient mapping of deep learning applications to mixed-precision hardware using model adaptation},
  year     = {2026},
  doi      = {10.1038/s41467-026-71071-1},
}

@InProceedings{Kao2022,
  author    = {Kao, Jui-I and others},
  booktitle = {Int. SoC Design Conf.},
  title     = {Precision-aware workload distribution and dataflow for a hybrid digital-{CIM} deep {CNN} accelerator},
  year      = {2022},
  doi       = {10.1109/ISOCC56007.2022.10031486},
}

@Article{Bai2024,
  author   = {Bai, Jinyu and Sun, Sifan and Zhao, Weisheng and Kang, Wang},
  journal  = {IEEE Trans. Comput.-Aided Design Integr. Circuits Syst.},
  title    = {{CIMQ}: A hardware-efficient quantization framework for computing-in-memory-based neural network accelerators},
  year     = {2024},
  volume   = {43},
  number   = {1},
  doi      = {10.1109/TCAD.2023.3298705},
}

@InProceedings{Peng2019,
  author    = {Peng, Xiaochen and others},
  booktitle = {IEEE Int. Electron Devices Meeting},
  title     = {{DNN+NeuroSim}: An end-to-end benchmarking framework for compute-in-memory accelerators with versatile device technologies},
  year      = {2019},
  doi       = {10.1109/IEDM19573.2019.8993491},
}

@InProceedings{Chen2024,
  author    = {Chen, Yung-Chin and others},
  booktitle = {Asia and South Pacific Design Automation Conf.},
  title     = {{OSA-HCIM}: On-the-fly saliency-aware hybrid {SRAM} {CIM} with dynamic precision configuration},
  year      = {2024},
  publisher = {IEEE Press},
  doi       = {10.1109/ASP-DAC58780.2024.10473966},
}

@InProceedings{Behnam2024,
  author    = {Behnam, Payman and others},
  booktitle = {Int. Parallel and Distributed Processing Symp.},
  title     = {Harmonica: Hybrid accelerator to overcome imperfections of mixed-signal {DNN} accelerators},
  year      = {2024},
  doi       = {10.1109/IPDPS57955.2024.00061},
}

@Article{Shafiee2016,
  author   = {Shafiee, Ali and Nag, Anirban and Muralimanohar, Naveen and others},
  journal  = {SIGARCH Comput. Archit. News},
  title    = {{ISAAC}: A convolutional neural network accelerator with in-situ analog arithmetic in crossbars},
  year     = {2016},
  volume   = {44},
  number   = {3},
  doi      = {10.1145/3007787.3001139},
}

@Article{Dash2022,
  author   = {Dash, Saurabh and others},
  journal  = {IEEE Trans. Comput.-Aided Design Integr. Circuits Syst.},
  title    = {Robust processing-in-memory with multibit {ReRAM} using {Hessian}-driven mixed-precision computation},
  year     = {2022},
  volume   = {41},
  number   = {4},
  doi      = {10.1109/TCAD.2021.3078408},
}

@Article{Risso2025,
  author   = {Risso, Matteo and Burrello, Alessio and Pagliari, Daniele Jahier},
  journal  = {IEEE Trans. Comput.-Aided Design Integr. Circuits Syst.},
  title    = {Optimizing {DNN} inference on multi-accelerator {SoCs} at training-time},
  year     = {2025},
  doi      = {10.1109/TCAD.2025.3543715},
}

@Article{Lammie2025,
  author   = {Lammie, Corey and others},
  journal  = {IEEE Trans. Emerging Topics Comput.},
  title    = {{LionHeart}: A layer-based mapping framework for heterogeneous systems with analog in-memory computing tiles},
  year     = {2025},
  doi      = {10.1109/TETC.2025.3546128},
}

@Article{Klein2023,
  author   = {Klein, Joshua and others},
  journal  = {IEEE Trans. Comput.},
  title    = {{ALPINE}: Analog in-memory acceleration with tight processor integration for deep learning},
  year     = {2023},
  volume   = {72},
  number   = {7},
  doi      = {10.1109/TC.2022.3230285},
}

@Article{Chen2025,
  author   = {Chen, Yizhe and others},
  journal  = {Moore and More},
  title    = {A reconfigurable heterogeneous in-memory computing architecture for variable precision computation: a software-hardware co-design approach},
  year     = {2025},
  volume   = {2},
  number   = {1},
  doi      = {10.1007/s44275-025-00028-1},
}

@Misc{3DCiM,
  author       = {{IBM}},
  title        = {{3D-CiM-LLM-Inference-Simulator}},
  year         = {2026},
  howpublished = {GitHub, \url{https://github.com/IBM/3D-CiM-LLM-Inference-Simulator}},
}

@Article{AIHWKit,
  author  = {Le~Gallo, Manuel and others},
  title   = {Using the {IBM} analog in-memory hardware acceleration kit for neural network training and inference},
  year    = {2023},
  journal = {APL Machine Learning},
  volume  = {1},
  doi     = {10.1063/5.0168089},
}

@InProceedings{Andrulis2024,
  author    = {Tanner Andrulis and Joel S. Emer and Vivienne Sze},
  booktitle = {{IEEE} Int. Symp. on Performance Analysis of Systems and Software},
  date      = {2024},
  title     = {{CiMLoop}: {A} Flexible, Accurate, and Fast Compute-In-Memory Modeling Tool},
  doi       = {10.1109/ISPASS61541.2024.00012},
  publisher = {{IEEE}},
}

\end{document}